\documentclass[aps,prc,preprint,amsmath,amssymb,showpacs,preprintnumbers,superscriptaddress]{revtex4}%
\usepackage{multirow}
\usepackage{rotating}
\usepackage{booktabs}
\usepackage{color,graphicx}
\usepackage{amsmath}
\usepackage{dcolumn}
\usepackage{CJK}
\usepackage{bm}
\usepackage{soul}
\usepackage{enumerate}
\usepackage[dvipdfm,bookmarks=true,colorlinks,
citecolor=blue,linkcolor=blue,hypertex,            breaklinks=true]{hyperref}
\usepackage{amsfonts}
\usepackage{amssymb}
\usepackage{graphicx}
\providecommand{\U}[1]{\protect\rule{.1in}{.1in}}

\begin{document}
\begin{CJK*}{GBK}{song}
\title{Shell-model-like approach based on cranking covariant density functional theory with a
separable pairing force}
\author{B.~W.~Xiong}
\affiliation{State Key Laboratory of Nuclear Physics and Technology, School of Physics,
Peking University, Beijing 100871, China}
\date{\today}
\begin{abstract}
The shell-model-like approach (SLAP) based on cranking covariant density functional theory
(CDFT) with a separable pairing force is developed. The developed cranking CDFT-SLAP with
separable pairing force is applied to investigate the rotational spectra in~$^{60}$Fe,
including the positive-parity yrast band and two negative-parity signature partner bands,
in comparison with the cranking CDFT-SLAP with monopole pairing force calculations.
Excellent agreement with the available data is achieved.
\\

\end{abstract}

\date{\today}


\maketitle
\section{Introduction}\label{sec1}

The study of nuclear rotation has been at the forefront of nuclear structure physics for
several decades. Many exciting phenomena have been discovered, such as backbending
\cite{johnson1971evidence,stephens1972coriolis}, superdeformed rotation
\cite{twin1986observation}, magnetic rotation \cite{Frauendorf1994Proceedings,
Frauendorf1997Tilted2}, antimagnetic rotation \cite{frauendorf2001spontaneous,
hubel2005magnetic}, chiral rotation \cite{frauendorf1997tilted,meng2010open,
meng2011chirality, meng2014chirality,meng2016nuclear,xiong2019nuclear}, and wobbling
motion \cite{bohr1975mottelson,timar2019experimental}. To achieve a unified description
of these phenomena is a challenge for the nuclear models.

The covariant density functional theory (CDFT) takes Lorentz symmetry into account in a
self-consistent way and has received wide attention due to its successful description of
a large number of nuclear phenomena in stable as well as exotic nuclei
\cite{ring1996relativistic,meng2006relativistic,meng2015halos,liang2015hidden,
meng2016relativistic}. For nuclear rotation, in particular, CDFT provides a consistent
description of currents and time-odd fields, and the included nuclear magnetism plays an
important role in one-dimensional principal axis cranking (PAC) \cite{koepf1990relativistic},
two-dimensional planar tilted axis cranking (TAC) \cite{peng2008covariant,zhao2011novel,
meng2013progress}, and three-dimensional aplanar TAC \cite{madokoro2000relativistic,
zhao2017multiple}. With these versions of cranking CDFT, novel rotational phenomena
including the magnetic rotational bands \cite{madokoro2000relativistic,peng2008covariant,
zhao2011novel,yu2012magnetic}, antimagnetic rotational bands \cite{zhao2011antimagnetic,
zhao2012covariant}, linear cluster structure \cite{zhao2015rod, ren2019stability}, chiral
rotational bands \cite{meng2016nuclear}, and multiple chiral doublets \cite{meng2006possible,
peng2008search,yao2009candidate,li2011multiple,Qi2013Possible,zhao2017multiple} have been
investigated successfully.

In these versions of cranking CDFT, pairing correlations are usually neglected or treated
by the Bardeen-Cooper-Schrieffer (BCS) approximation or Bogoliubov transformation
\cite{ring2004nuclear}. To overcome the problems of particle number non-conservation
\cite{zeng1983particle}, the blocking effect \cite{rowe1970nuclear}, and the pairing
collapse with rotation \cite{mottelson1960effect}, the shell-model-like approach (SLAP)
\cite{zeng1983particle,meng2006shell} based on the cranking CDFT has been developed to 
treat pairing correlations with exact particle number conservation \cite{shi2018shell}.

Originally referred to as the particle-number-conserving (PNC) method \cite{zeng1983particle},
SLAP treats pairing correlations and blocking effects exactly by diagonalizing the many-body
Hamiltonian in a many-particle configuration (MPC) space with conserved particle number.
Based on the cranking Nilsson model, extensive applications for the odd-even differences in
moments of inertia \cite{zeng1994blocking},~identical bands \cite{liu2002microscopic,he200513},
nuclear pairing phase transition \cite{wu2011nuclear}, antimagnetic rotation
\cite{zhang2013nuclear,zhang2016effects}, and high-$K$ rotational bands in the rare-earth \cite{liu2004particle,zhang2009particle,Zhang2009Particle,li2013particle,li2016rotational,
zhang2016particle,zhang2013rotational}, and actinide \cite{he2009influence,zhang2011particle,
zhang2012systematic} nuclei, have been performed.

Based on the CDFT, the SLAP has been first adopted to study the ground-state properties and
low-lying excited states for Ne isotopes \cite{meng2006shell}. The self-consistency is
achieved by iterating the occupation probabilities from SLAP back to the densities and
currents in CDFT. Along this line, the extension to include the temperature has been used
to study the heat capacity \cite{liu2015thermodynamics}. The SLAP has also been combined
with deformed Woods-Saxon potential \cite{molique1997fock,fu2013configuration} and Skyrme
density functional \cite{pillet2002pairing,liang2015configuration}.

The cranking CDFT-SLAP with monopole pairing force has been developed to study the band
crossing and shape evolution in $^{60}$Fe \cite{shi2018shell} and the antimagnetic rotation
band in $^{101}$Pd \cite{liu2019shell}. A separable version of the Gogny pairing force,
which can be represented as a sum of a finite number of separable terms in the harmonic
oscillator basis, was introduced by Tian $et~al$. \cite{tian2009finite}. The separable
pairing force is finite range and, thus, the problem of an ultraviolet divergence can be
avoided. Meanwhile, due to its separable form, it requires less computational time as
compared to other finite range pairing forces. The separable pairing force has been
implemented in the TAC-CDFT and applied to the yrast band in $^{109}$Ag \cite{wang2017yrast}
as well as the magnetic rotational bands in $^{198}$Pb and $^{199}$Pb \cite{wang2018magnetic}.

In the present work, the SLAP based on the cranking CDFT with a separable pairing force
is developed. In Sec.~\ref{sec2}, the theoretical framework of the cranking CDFT-SLAP
with  separable pairing force is briefly presented. The numerical details are given in
Sec.~\ref{sec3}. In Sec.~\ref{sec4}, the energy spectra, the pairing energies, and the
shape evolutions for the three rotational bands in $^{60}$Fe are calculated and compared
with the data available \cite{deacon2007yrast} as well as the results in Ref.
\cite{shi2018shell} given by the cranking CDFT-SLAP with monopole pairing force. Finally,
a short summary is given in Sec.~\ref{sec5}.
\section{Theoretical framework}\label{sec2}
\subsection{Cranking CDFT}

The starting point of the CDFT based on point-coupling interaction is a standard
effective Lagrangian density \cite{burvenich2002nuclear,nikvsic2008relativistic,
zhao2010new}. For rotating nucleus, one can transform the effective Lagrangian into a
rotating frame with a constant rotational frequency $\omega$ around a fixed direction
\cite{koepf1989relativistic,konig1993identical,kaneko1993three}. This gives rise to the
PAC-CDFT \cite{koepf1990relativistic}, where the cranking axis is one of the three
principal axes of a nucleus, or the TAC-CDFT with the cranking axis different from any
of the principal axes, including planar \cite{peng2008covariant,zhao2011novel,
meng2013progress} and aplanar rotation versions \cite{madokoro2000relativistic,
zhao2017multiple}.

From this rotating Lagrangian, the equation of motion for the nucleus can be derived as
\cite{meng2013progress,shi2018shell}
\begin{equation}\label{eq-dirac-1d}
   \hat h'_{\rm s.p.}\psi_\mu = (\hat h_{\rm s.p.}+\hat h_{\rm c})\psi_\mu
                              = \varepsilon_\mu\psi_\mu,
\end{equation}
with
\begin{align}
    \hat h_{\rm s.p.} = \bm \alpha\cdot(-i{\bm\nabla}-\bm{V})+\beta(m+S)+V^0,~~~~
    \hat h_{\rm c}    = -{\omega_x\hat{j}_x},
\end{align}
where $ \hat j_x =\bm \hat l_x+\frac{1}{2} \hat \Sigma_x$ is the $x$ component of the
total angular momentum of the nucleon spinors, and $\varepsilon_\mu$ represents the
single-particle Routhians. The relativistic scalar $S({\bm r})$ and vector $V^\mu(\bm r)$
fields are connected in a self-consistent way to the densities and currents
\cite{meng2013progress}.

The equation of motion (\ref{eq-dirac-1d}) can be solved by expanding the nucleon spinors
in a complete set of basis states. The three-dimensional harmonic oscillator (3DHO) bases
in Cartesian coordinates \cite{peng2008covariant,koepf1988has,dobaczewski1997solution,
yao2006time,nikvsic2009beyond} with good signature quantum number are adopted,
\begin{align} \label{3dho-nega}
 \Phi_{a+}({\bm r}, {\bm s})
    =\langle {\bm r}, {\bm s}|a\alpha=+\rangle
   &=\phi_{n_x}\phi_{n_y}\phi_{n_z}\frac{i^{n_y}}{\sqrt{2}}(-1)^{n_z+1}
     \left(\begin{array}{ccc}
      1\\
      (-1)^{n_y+n_z}
     \end{array}\right), \\
     \Phi_{a-}({\bm r}, {\bm s})
   =\langle {\bm r}, {\bm s}|a\alpha=-\rangle
  &=\phi_{n_x}\phi_{n_y}\phi_{n_z}\frac{i^{n_y}}{\sqrt{2}}
    \left(\begin{array}{ccc}
     1\\
     (-1)^{n_y+n_z+1}
    \end{array}\right),
    \label{3dho-posi}
\end{align}
which correspond to the eigenfunctions of the signature operation with the positive
$(\alpha=+1/2)$ and negative $(\alpha=-1/2)$ eigenvalues, respectively. The $n_x$, $n_y$,
and $n_z$ represent the harmonic oscillator quantum numbers in $x$, $y$, and $z$
directions, and $\phi_{n_x}$, $\phi_{n_y}$, and $\phi_{n_z}$ denote the corresponding
eigenstates.

By solving the Dirac equation (\ref{eq-dirac-1d}) with given signature $\alpha$
self-consistently, the single-particle Routhian $\varepsilon_{\mu\alpha}$ and the
corresponding eigenstate $\psi_{\mu\alpha}$ for each level $\mu$ can be obtained
\cite{meng2013progress,meng2016relativistic}.
\subsection{Cranking CDFT-SLAP}

The cranking CDFT-SLAP starts from a cranking many-body Hamiltonian including pairing
correlations
\begin{align}\label{hami-eq-crank}
   \hat H=\hat H'+\hat H_{\rm pair},
\end{align}
where $\hat H'=\sum\hat h'_{\rm s.p.}$ is the one-body Hamiltonian with
$\hat h'_{\rm s.p.}$ defined in Eq.~(\ref{eq-dirac-1d}). The pairing Hamiltonian
$\hat H_{\rm pair}$ is expressed as
\begin{equation}
\hat{H}_{\mathrm{pair}}=\frac{1}{2}\sum_{abcd}\langle ab|\hat{V}
                 _{\mathrm{pair}}|cd\rangle
                 \hat{\beta}_{a}^\dag \hat{\beta}_{b}^\dag
                 \hat{\beta}_{d} \hat{\beta}_{c},
\end{equation}
where the create and annihilate operators of the 3DHO bases are denoted by
$\hat{\beta}_{a}^\dag$,~$\hat{\beta}_{b}^\dag$,~$\hat{\beta}_{c}$, and
$\hat{\beta}_{d}$, respectively, and $\hat V_{\mathrm{pair}}$ is the separable
pairing force \cite{tian2009finite},
\begin{equation}
\label{pp-force}
\begin{split}
\hat V_{\mathrm{pair}}(\bm{r}_1,\bm{r}_2;\bm{r}_1^\prime,\bm{r}_2^\prime)
    &=G\delta \left(\bm{R}-\bm{R}^\prime \right)P(\bm{r})P(\bm{r}^\prime)
      \frac{1}{2}\left(1-P^\sigma \right).
\end{split}
\end{equation}
Here, $\bm{R}=\frac{1}{2}(\bm{r}_1+\bm{r}_2)$ and $\bm{r}=\bm{r}_1-\bm{r}_2$
denote the center-of-mass and the relative coordinates, respectively, and
$P(\bm{r})$ is the Gaussian function
\begin{equation}
P(\bm{r})=\frac{1}{(4\pi a^2)^{3/2}}\mathrm{e}^{-\frac{r^2}{4a^2}}.
\end{equation}
The projector $\frac{1}{2}(1-P^\sigma)$ allows only the states with the total spin
$S=0$. The two parameters $G$ and $a$ were determined in Ref. \cite{tian2009finite}
by fitting to the density dependence of pairing gaps at the Fermi surface for
nuclear matter obtained with the Gogny forces.

In the 3DHO bases~(\ref{3dho-nega})-(\ref{3dho-posi}), the one-body Hamiltonian
$\hat H'$ can be written as
\begin{equation}\label{eq-h'-base}
\hat H'=\sum_{ab,\alpha}\langle a|\hat h'_{\rm s.p.}|b\rangle\hat
\beta_{a\alpha}^\dagger \hat \beta_{b\alpha}.
\end{equation}
Accordingly, the pairing Hamiltonian $\hat H_{\mathrm{pair}}$ in the 3DHO bases
can be written as
\begin{equation}
\hat{H}_{\mathrm{pair}}=\frac{1}{2}\sum_{abcd}
    \langle ab|\hat{V}_{\mathrm{pair}}|cd\rangle
    \hat{\beta}_{a\alpha_1}^\dag \hat{\beta}_{b\alpha_2}^\dag
    \hat{\beta}_{d\alpha_4} \hat{\beta}_{c\alpha_3}.
\end{equation}

The idea of SLAP is to diagonalize the many-body Hamiltonian in a properly
truncated MPC space with exact particle number \cite{zeng1983particle}. In the
present work, the cranking many-body Hamiltonian (\ref{hami-eq-crank}) is
diagonalized in the MPC space constructed from the single-particle states in the
cranking CDFT. Except that the original monopole pairing force is replaced by the
present separable pairing force, the other formalisms are the same as those in 
Ref. \cite{shi2018shell}.

Diagonalizing the one-body Hamiltonian $\hat H'$ (\ref{eq-h'-base}) in the bases
$|a\alpha\rangle$ (\ref{3dho-nega})-(\ref{3dho-posi}), one can obtain the
single-particle Routhian $\varepsilon_{\mu\alpha}$ and the corresponding eigenstate
$|\mu\alpha\rangle$ for each level $\mu$ with the signature $\alpha$, namely,
\begin{align}
    \hat H' =\sum_{\mu\alpha} \varepsilon_{\mu\alpha}
              \hat b^\dag_{\mu\alpha}\hat b_{\mu\alpha},~~~~~
    |\mu\alpha\rangle = \sum_a C_{\mu a}(\alpha)|a\alpha\rangle.
\end{align}
From the real expansion coefficient $C_{\mu a}(\alpha)$, the transformation between
the operators $\hat b^\dag_{\mu\alpha}$ and $\hat\beta^\dag_{a\alpha}$ can be
expressed as
\begin{align}\label{eq-state-base}
    \hat b^\dag_{\mu\alpha}=\sum_a C_{\mu a}(\alpha)\hat \beta^\dag_{a\alpha}, ~~~~~
    \hat \beta^\dag_{a\alpha}=\sum_\mu C_{\mu a}(\alpha)\hat b^\dag_{\mu\alpha}.
\end{align}

In the $|\mu\alpha\rangle$ basis, the pairing Hamiltonian $\hat H_{\rm pair}$ can
be written as
\begin{equation}\label{Hpair-sp-basis}
\begin{split}
 \hat{H}_{\mathrm{pair}}
  =&\frac{1}{2}\sum_{abcd}
   \langle ab|\hat{V}_{\rm{pair}}|cd\rangle\\
   &\times
   \big(\sum_{\mu_1}C_{\mu_1 a}(\alpha_1)\hat b^\dag_{\mu_1\alpha_1}\big)
   \big(\sum_{\mu_2}C_{\mu_2b}(\alpha_2)\hat b^\dag_{\mu_2\alpha_2}\big)
   \big(\sum_{\mu_4}C_{\mu_4d}(\alpha_4)\hat b_{\mu_4\alpha_4}\big)
   \big(\sum_{\mu_3}C_{\mu_3c}(\alpha_3)\hat b_{\mu_3\alpha_3}\big)\\
  =&\frac{1}{2}
   \sum_{abcd}\sum_{\mu_1\mu_2\mu_3\mu_4}
   \langle ab|\hat{V}_{\rm{pair}}|cd\rangle
   ~C_{\mu_1a}(\alpha_1)C_{\mu_2b}(\alpha_2)
   C_{\mu_4d}(\alpha_4)
   C_{\mu_3c}(\alpha_3)
   \hat b^\dag_{\mu_1\alpha_1}\hat b^\dag_{\mu_2\alpha_2}
   \hat b_{\mu_4\alpha_4}\hat b_{\mu_3\alpha_3}.
\end{split}
\end{equation}

Based on the single-particle Routhian $\varepsilon_{\mu\alpha}$ and the
corresponding eigenstate $|\mu\alpha\rangle$ (briefly denoted by $|\mu\rangle$),
the MPC $|i\rangle$ for an $n$-particle system can be constructed as
\cite{zeng1994reduction}
\begin{align}
   |i\rangle= |\mu_1\mu_2\cdots\mu_n\rangle
            = \hat b^\dag_{\mu_1}\hat b^\dag_{\mu_2}\cdots
            \hat b^\dag_{\mu_n}|0\rangle.
\end{align}
The parity $\pi$, signature $\alpha$, and the corresponding configuration
energy for each MPC are determined by the occupied single-particle states.

The eigenstates for the cranking many-body Hamiltonian are obtained by
diagonalization in the MPC space,
\begin{align}
    |\Psi\rangle = \sum_i C_i |i\rangle,
\end{align}
where $C_i$ are the expanding coefficients.

The occupation probability $n_\mu$ for state $\mu$ is defined as
\begin{align}\label{eq-occupy}
    n_\mu=\sum_i |C_i|^2P_{i\mu},~~~~
    P_{i\mu}=\left\{
\begin{array}{cl}
    1, &~~~~|i\rangle~{\rm contains~|\mu\rangle},\\
    0, &~~~~{\rm otherwise}.
\end{array}
\right.
\end{align}
The occupation probabilities will be iterated back into the densities and
currents in the relativistic scalar and vector fields to achieve
self-consistency~\cite{meng2006shell}.

It is noted that, for the total energy in CDFT, the pairing energy due to
the pairing correlations should be taken into account,
$E_{\rm pair} = \langle\Psi|\hat H_{\rm pair}|\Psi\rangle$.

For each rotational frequency $\omega_x$, the expectation value of the
angular momentum $J_x$ in the intrinsic frame is given by
\begin{equation}
J_x=\langle\Psi|\hat{J}_x|\Psi\rangle=\sum_iC_i^2
\langle i|\hat{J}_x|i\rangle+\sum_{i,j}C_iC_j\langle i|\hat{J}_x|j\rangle,
\end{equation}
and by means of the semiclassical cranking condition
\begin{equation}
J_x=\langle\Psi|\hat{J}_x|\Psi\rangle\equiv \sqrt{I(I+1)},
\end{equation}
one can relate the rotational frequency $\omega_x$ to the angular momentum
quantum number $I$ in the rotational band.
\section{Numerical details}\label{sec3}

In the present cranking CDFT-SLAP calculations for $^{60}$Fe, the point-coupling
density functional PC-PK1 \cite{zhao2010new} is used in the particle-hole channel,
and the separable pairing \cite{tian2009finite} is adopted in the particle-particle
channel, respectively.
\begin{figure*}[h!]
  \centering
  \includegraphics[scale=0.215,angle=0]{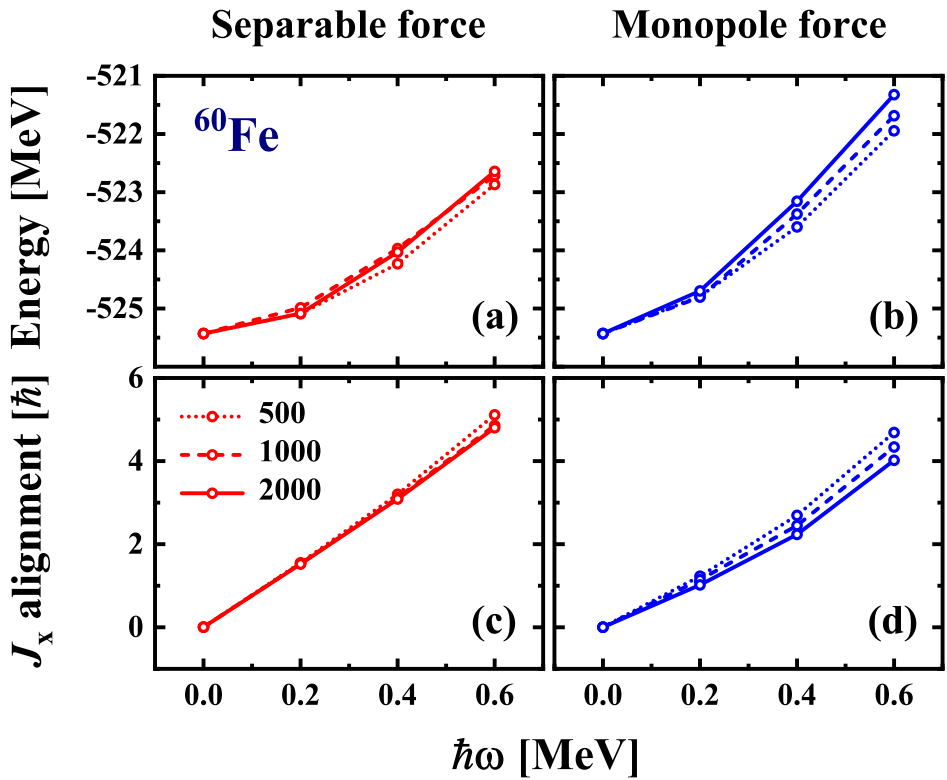}
  \caption{(Color online) The total energies (upper panels) and the alignments along
  the rotational axis (lower panels) as functions of the rotational frequency in
  $^{60}$Fe calculated by the cranking CDFT-SLAP with separable pairing force (left
  panels) and monopole pairing force (right panels), respectively. The separable
  pairing force with $G=-728$~MeV~fm$^3$ and $a=0.644$~fm \cite{tian2009finite}, and
  the monopole pairing force with $G_n=G_p=0.8$~MeV \cite{shi2018shell} are used in
  the particle-particle channel. For both neutron and proton, the dimensions of the
  MPC space are chosen as 500~(dotted lines), 1000~(dashed lines), and 2000 (solid
  lines), respectively. The energy at $\hbar\omega=0$ MeV calculated by separable
  pairing force is taken as reference.}
\label{fe60-convergence-check}
\end{figure*}

Similar to Ref. \cite{shi2018shell}, by switching off pairing correlations, the
validity of the cranking CDFT-SLAP with separable pairing force is checked against
the TAC-CDFT calculation \cite{zhao2011novel}. Here the total energies and the
alignments along the rotational axis as functions of the rotational frequency in
$^{60}$Fe calculated by the cranking CDFT-SLAP, are compared with the TAC-CDFT
calculations with tilted angle $\theta=0^\circ$ \cite{zhao2011novel}. Satisfactory
agreement is found with the differences less than 10$^{-4}$ MeV for the total energy
and 10$^{-4}\hbar$ for the alignment.

The convergence with respect to the major oscillator shells $N_f$ has been checked.
By increasing $N_f$ from 12 to 14, the changes of the total energy and alignment for
$\hbar\omega=0.2$~MeV are only 0.004$\%$ and 0.760$\%$, respectively. Thus $N_f=12$
are used in the present calculations.

The convergence with respect to the dimension of the MPC space has also been checked.
The calculated total energies and alignments in $^{60}$Fe by the cranking CDFT-SLAP
with separable pairing force are shown in Fig.~\ref{fe60-convergence-check}, compared
with the results from the cranking CDFT-SLAP with monopole pairing force. For the
separable pairing force, the tendency of convergence can be clearly seen. By increasing
the dimensions of the MPC space from 1000 to 2000, the changes of the total energy and
alignment for $\hbar\omega=0.2$~MeV are 0.151$\%$ and 0.667$\%$, respectively. In the
following calculations, the dimensions of the MPC space are 1000 for both neutron and
proton. For the monopole pairing force, there are no tendency of convergence with
respect to the dimension of the MPC space for both the total energy and alignment. As
was pointed out in Ref. \cite{shi2018shell}, the effective pairing strengths have to be
changed when changing the dimension of the MPC space in this case.
\section{Results and discussion}\label{sec4}

Three rotational bands in the nucleus $^{60}$Fe, including the positive-parity yrast
band (labeled as band A) and two negative-parity signature partner bands (labeled as
bands B and C), were observed in Ref. \cite{deacon2007yrast}. The cranking CDFT-SLAP
with monopole pairing force has been applied to investigate these three bands
\cite{shi2018shell}. In the following, the cranking CDFT-SLAP with separable pairing
force will be used to calculate these three bands and compared with the data and the
results in Ref.~\cite{shi2018shell}.
\subsection{Energy spectra}

\begin{figure*}[h!]
  \centering
  \includegraphics[scale=0.242,angle=0]{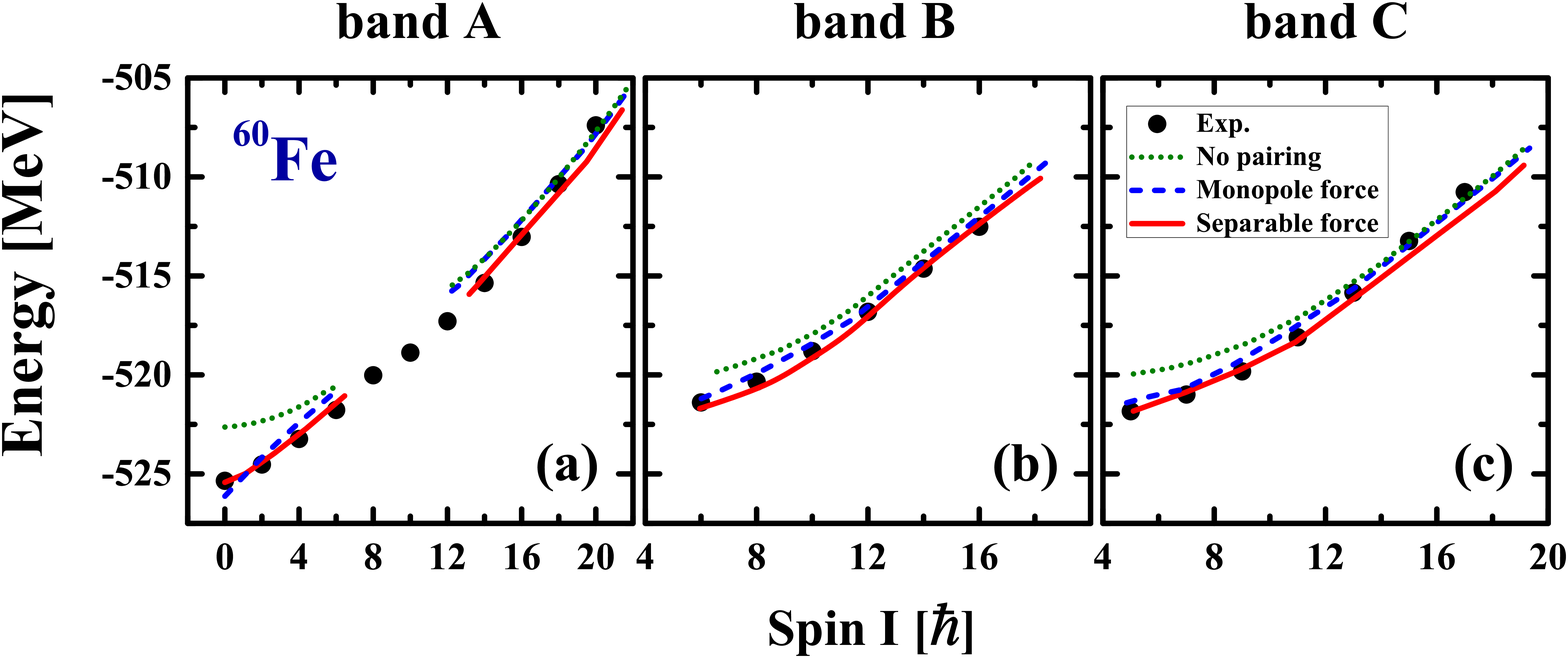}
  \caption{(Color online)
  The total energies for the positive-parity band A (left panel), negative-parity
  signature partner bands B (middle panel) and C (right panel) in $^{60}$Fe as
  functions of the spin calculated by the cranking CDFT-SLAP with separable pairing
  force (solid lines), in comparison with the available data \cite{deacon2007yrast}
  (solid dots) and the results calculated by the cranking CDFT-SLAP with monopole
  pairing force (dashed lines) and without pairing (dotted lines) from Ref.
  \cite{shi2018shell}.}
\label{fe60-energy-spectra}
\end{figure*}

In Fig.~\ref{fe60-energy-spectra}, the calculated total energies for the positive-parity
band A and negative-parity signature partner bands B and C in $^{60}$Fe are shown in
comparison with the data \cite{deacon2007yrast} and the results calculated by the
cranking CDFT-SLAP with monopole pairing force and without pairing from Ref.
\cite{shi2018shell}.

For band A, it is found that the cranking CDFT-SLAP with separable pairing force
provides a successful description of the energy spectra, and are comparable with the
cranking CDFT-SLAP with monopole pairing force. Both of them have a significant
improvement on the results without pairing, in particular for the low-spin regions.
There are sudden discontinuities in experimental energy sequence and intensities of
the transitions at $I=8\hbar$, indicating a structural change and band crossing
\cite{deacon2007yrast}. Theoretically, all the three calculations can give this sudden
change in the band structure. The band crossing obtained by the cranking CDFT-SLAP
with separable pairing force occurs a bit later.

For band B, one can see that a better agreement with the data is obtained in the cranking
CDFT-SLAP with separable pairing force than the cranking CDFT-SLAP with monopole pairing
force, especially for the bandhead. Similar conclusion holds for band C.
\subsection{$I-\omega$ relations}

\begin{figure*}[h!]
  \centering
  \includegraphics[scale=0.242,angle=0]{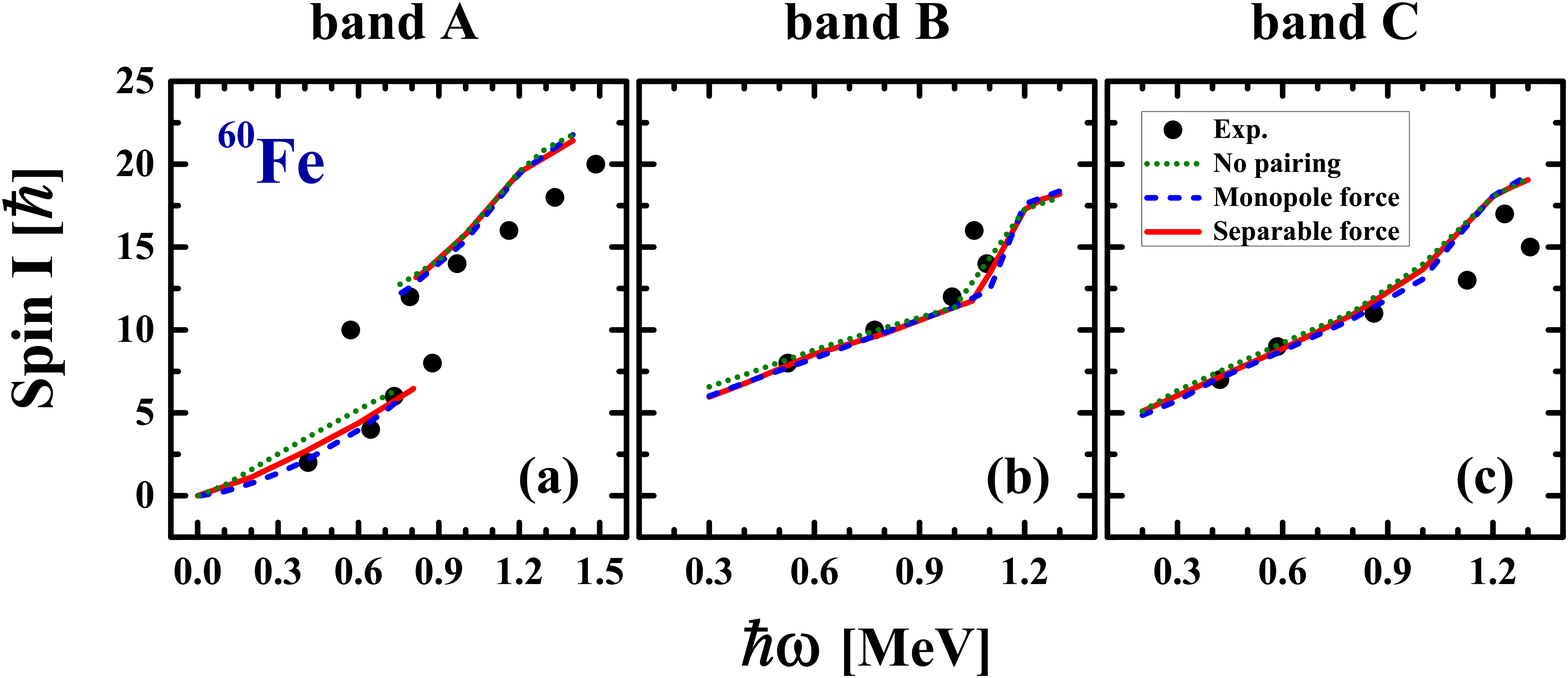}
  \caption{(Color online)
  The angular momenta for the positive-parity band A (left panel), negative-parity
  signature partner bands B (middle panel) and C (right panel) in $^{60}$Fe as functions
  of the rotational frequency calculated by the cranking CDFT-SLAP with separable pairing
  force (solid lines), in comparison with the available data \cite{deacon2007yrast} (solid
  dots), and the results calculated by the cranking CDFT-SLAP with monopole pairing force
  (dashed lines) and without pairing (dotted lines) from Ref. \cite{shi2018shell}.}
\label{fe60-I-omega}
\end{figure*}

In Fig.~\ref{fe60-I-omega}, the calculated angular momenta as functions of the rotational
frequency are shown in comparison with the data \cite{deacon2007yrast}, and the results
calculated by the cranking CDFT-SLAP with monopole pairing force as well as without
pairing from Ref. \cite{shi2018shell} for bands A, B, and C.

For band A, it is found that the inclusion of pairing correlations brings an improvement to
the description of the $I\sim \hbar\omega$ relation, and both results given by the cranking
CDFT-SLAP with separable pairing force and monopole pairing force agree well with the data.
Experimentally, the $I\sim \hbar\omega$ relation shows an irregularity at spin $I=8\hbar$.
As discussed in Ref. \cite{shi2018shell}, this corresponds to the sudden change of the
configuration. The band crossing frequency obtained from the cranking CDFT-SLAP with separable
pairing force is $\hbar\omega \sim  0.85$ MeV, which is a little larger than that from the
cranking CDFT-SLAP calculations with monopole pairing force and without pairing
($\hbar\omega\sim 0.75$ MeV) \cite{shi2018shell}.

For band B, the cranking CDFT-SLAP calculations with separable pairing force and monopole
pairing force give very similar results. Both of them reproduce the experimental band
crossing at $\hbar\omega=1.1$~MeV well \cite{deacon2007yrast}. For band C, similar
conclusion with band B can be drawn, only that the predicted band crossing is somewhat
earlier than the experimental one.
\subsection{Pairing energies}

\begin{figure*}[h!]
  \centering
  \includegraphics[scale=0.24,angle=0]{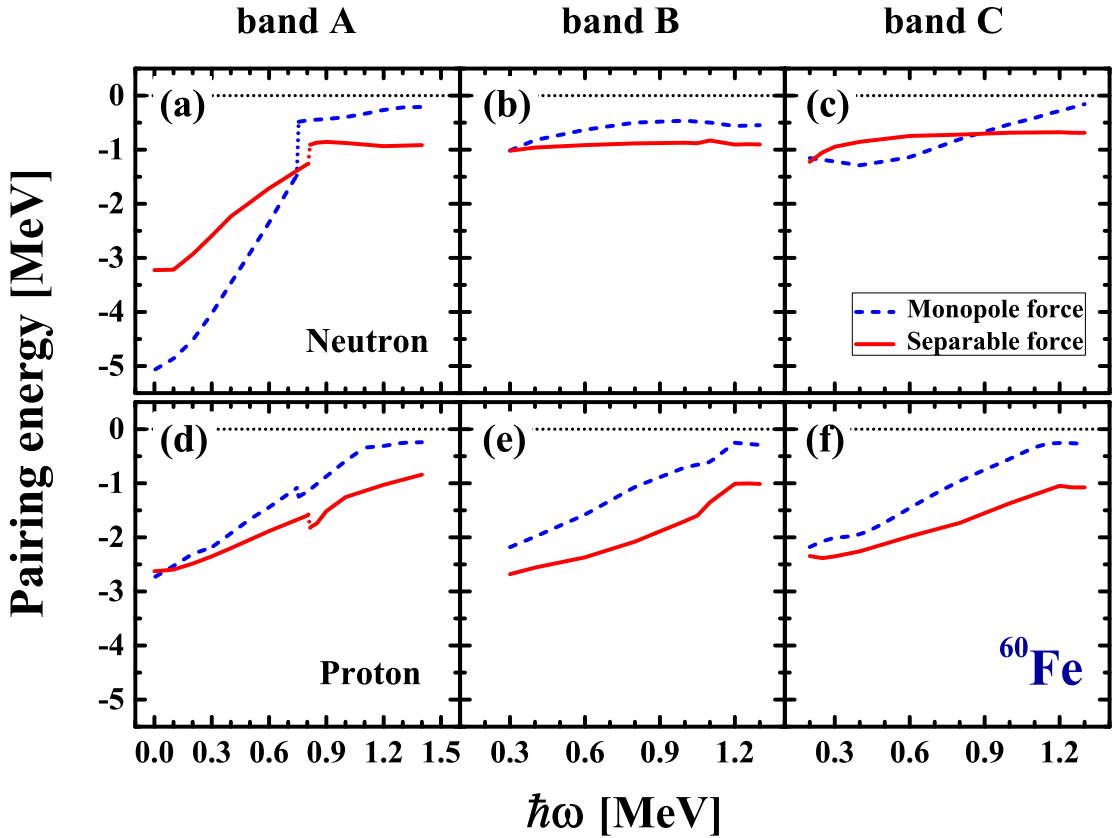}
  \caption{(Color online)
  The neutron and proton pairing energies for the positive-parity band A (left panel),
  negative-parity signature partner bands B (middle panel), and C (right panel) in
  $^{60}$Fe as functions of the rotational frequency calculated by the cranking
  CDFT-SLAP with separable pairing force (solid lines), in comparison with the results
  calculated by the cranking CDFT-SLAP with monopole pairing force (dashed lines) from
  Ref. \cite{shi2018shell}.}
\label{fe60-pairing-energy}
\end{figure*}

One of the advantages of SLAP is that the pairing correlations are treated exactly and the
particle number is conserved, thus there is no sharp pairing collapse in the calculations.
Fig.~\ref{fe60-pairing-energy} shows the neutron and proton pairing energies as functions
of the rotational frequency, in comparison with the results calculated by the cranking
CDFT-SLAP with monopole pairing force from Ref. \cite{shi2018shell} for bands A, B, and C.

As seen in Fig.~\ref{fe60-pairing-energy}, both of the neutron and proton pairing energies
from both the cranking CDFT-SLAP calculations with separable pairing force and monopole
pairing force decrease with the rotational frequency. There is no sharp pairing collapse
but rather a more continuous transition as rotational frequency is increased. However,
there are evident differences in quantity between them.

For band A, before band crossing, the neutron pairing energy obtained from the cranking 
CDFT-SLAP with separable pairing force is smaller than that from monopole pairing force, 
and decreases with a smaller slope with the rotational frequency. After band crossing, 
the neutron pairing energy from monopole pairing force drops drastically to almost zero, 
while that from separable pairing force stays at around 1~MeV. The reason for this can be 
understood. The monopole pairing force only takes into account Cooper pairs coupled to 
angular momentum $J=0$. In contrast, the separable pairing force takes into account 
correlations not only in pairs with $J=0$, but also in pairs with higher angular momentum 
\cite{tian2009finite}.
\subsection{Shape evolutions}

\begin{figure*}[h!]
  \centering
  \includegraphics[scale=0.24,angle=0]{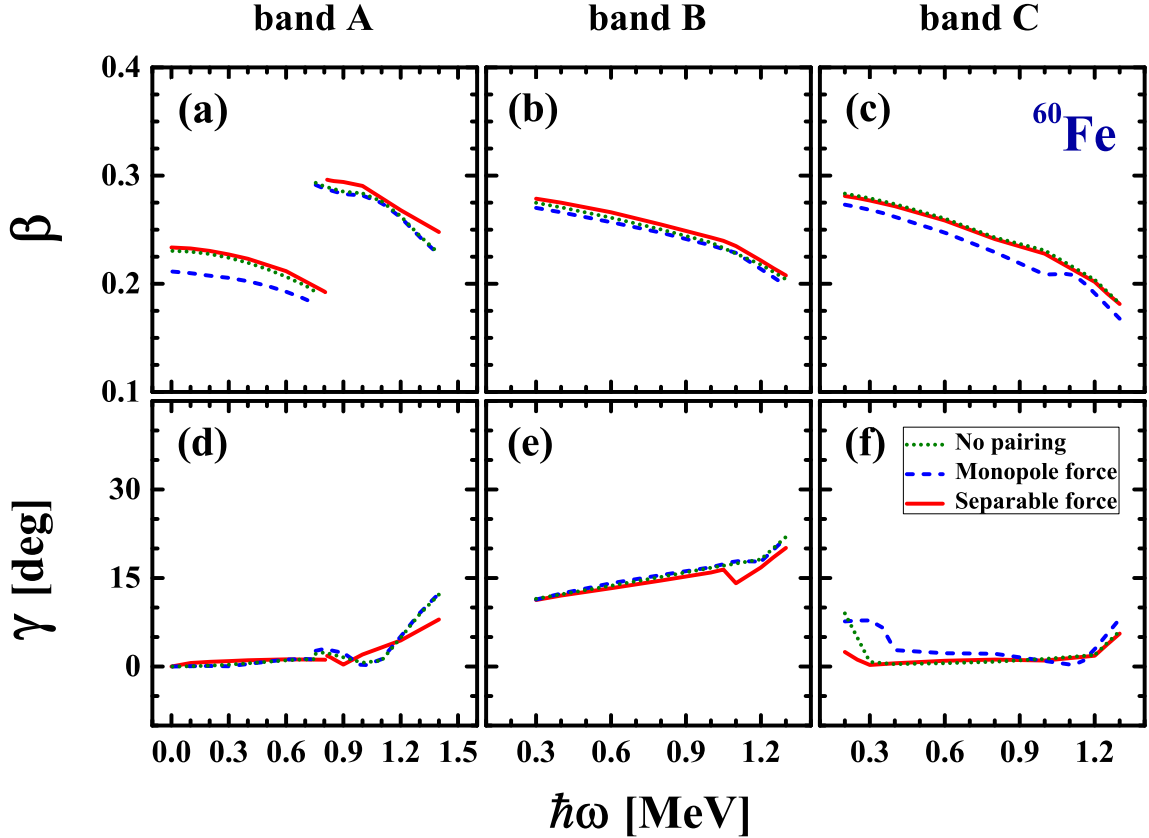}
  \caption{(Color online)
  The quadrupole deformation parameters $\beta$ and $\gamma$ for the positive-parity band A
  (left panel), negative-parity signature partner bands B (middle panel) and C (right panel)
  in $^{60}$Fe as functions of the rotational frequency calculated by the cranking CDFT-SLAP
  with separable pairing force (solid lines), in comparison with the results calculated by
  the cranking CDFT-SLAP with monopole pairing force (dashed lines) and without pairing
  (dotted lines) from Ref. \cite{shi2018shell}.}
\label{fe60-deformation}
\end{figure*}
In the CDFT calculation, the nuclear shape is obtained self-consistently. The shape evolutions
with the rotational frequency for the three bands in $^{60}$Fe have been investigated in Ref.
\cite{shi2018shell}. It was found that in general the deformation parameters $\beta$ for bands
A, B, and C decrease with the rotational frequency. For band A, the deformation jumps from
$\beta\approx$ 0.19 to $\beta\approx$ 0.29 around the band crossing. In comparison with its
signature partner band C, band B exhibits appreciable triaxial deformation \cite{shi2018shell}.

Figure~\ref{fe60-deformation} shows the evolutions of the quadrupole deformation parameters
$\beta$ and $\gamma$ with the rotational frequency obtained from the cranking CDFT-SLAP with
separable pairing force, in comparison with the corresponding results from monopole pairing
force and without pairing. Generally speaking, the results from these three theoretical
calculations are very similar. Therefore, the characteristics of the shape evolutions obtained
in Ref. \cite{shi2018shell} hold in the cranking CDFT-SLAP with separable pairing force 
calculations. It is noted that, the $\beta$ obtained from separable pairing force is slightly
larger than that from monopole pairing force. This owes to two possible reasons. On the one
hand, the neutron pairing energy obtained from separable pairing force is significantly smaller
than that from monopole pairing force near the bandhead in band A, in corresponding with $\beta$
obtained from former calculation is about 0.02 larger than that from latter calculation during
this region. On the other hand, the separable pairing force takes into account correlations in
pairs with higher angular momentum, which may lead to a slightly larger $\beta$ than that from
monopole pairing force in bands B, C and in the region after band crossing in band A although
the pairing energy from former calculation is larger.
\section{Summary}\label{sec5}

In summary, a finite range  separable pairing force is implemented in the shell-model-like
approach based on the cranking covariant density functional theory. This method has been applied
to investigate the rotational spectra observed in $^{60}$Fe, including the positive-parity band
A and negative-parity signature partner bands B and C, in comparison with the cranking CDFT-SLAP
with monopole pairing force calculations. The examination of the convergence with respect to the
MPC dimension shows that the calculation with separable pairing force can give better convergence
than that with monopole pairing force. Excellent agreement with the available data is achieved.
Furthermore, the pairing energies obtained from the cranking CDFT-SLAP with separable pairing
force and monopole pairing force show evident differences in quantity that in general the pairing
energy from the former calculation decreases slower with the rotational frequency than the latter
one. It may be due to separable pairing force takes into account correlations not only in pairs
with $J=0$, but also in pairs with higher angular momentum. This could also be the reason why the
quadrupole deformation $\beta$ obtained from separable pairing force is slightly larger than that
from monopole pairing force although the pairing energy from former calculation is larger.

\begin{acknowledgments}

The author is indebted to Prof. J. Meng for constructive guidances and valuable suggestions. The
author thanks Y. K. Wang and S. Q. Zhang for helpful discussions and careful readings of the
manuscript. Fruitful discussions with F. Q. Chen, Q. B. Chen, L. Liu, P. Ring, Z. Shi, Z. H. Zhang,
and P. W. Zhao are very much appreciated. This work was partly supported by the National Key
R$\&$D Program of China (Contracts No. 2017YFE0116700 and No. 2018YFA0404400) and the National
Natural Science Foundation of China (NSFC) under Grants No. 11335002, No. 11875075, and No.
11621131001.
\end{acknowledgments}

\begin{appendix}
\section{CALCULATION OF PAIRING MATRIX ELEMENTS}\label{AppendA}

The harmonic oscillator bases one uses to solve the equation of motion (\ref{eq-dirac-1d}) read
\begin{align}
  |n_xn_yn_z;\alpha&=+\rangle=|n_xn_yn_z\rangle\frac{i^{n_y}}{\sqrt{2}}(-1)^{
    n_z+1}\left[ |\uparrow\rangle+(-1)^{n_y+n_z}|\downarrow
    \rangle\right],\\
  |n_xn_yn_z;\alpha&=-\rangle=|n_xn_yn_z\rangle\frac{i^{n_y}}{\sqrt{2}}\left[
    |\uparrow\rangle+(-1)^{n_y+n_z+1}|\downarrow
    \rangle\right].
\end{align}
Here, $|n_xn_yn_z\rangle$ is the harmonic oscillator wave function in Cartesian coordinates,
and $n_x,~n_y,~n_z$ are the corresponding quantum numbers. The labels $\alpha=+$ and $\alpha=-$
represent the states with positive and negative signature, respectively, and for simplicity
they are respectively abbreviated below as $|a\rangle$ and $|\bar{b}\rangle$.

Based on these harmonic oscillator bases, the antisymmetric pairing matrix elements
$\langle ab|\hat{V}_{\mathrm{pair}}|cd\rangle_a$ in Eq. (\ref{Hpair-sp-basis}) can be
calculated, where the separable pairing force $\hat{V}_{\mathrm{pair}}$ (\ref{pp-force}) can
be written as
\begin{equation}
\begin{split}
  \hat V_{\mathrm{pair}}(\bm{r}_1,\bm{r}_2;\bm{r}_1^\prime,\bm{r}_2^\prime)
    &=G\delta \left(\bm{R}-\bm{R}^\prime \right)P(\bm{r})P(\bm{r}^\prime)
      \frac{1}{2}\left(1-P^\sigma \right)\\
    &\equiv W(\bm{r}_1,\bm{r}_2;\bm{r}_1^\prime,\bm{r}_2^\prime)\frac{1}{2}(1-P^\sigma).
\end{split}
\end{equation}

There are four types of such matrix elements, i.e., $\langle a\bar{b}|\hat{V}_{\mathrm{pair}}
|c\bar{d}\rangle_a$, $\langle ab|\hat{V}_{\mathrm{pair}}|cd\rangle_a$, $\langle ab|\hat{V}_
{\mathrm{pair}}|\bar{c}\bar{d}\rangle_a$,~and $\langle \bar{a}\bar{b}|\hat{V}_{\mathrm{pair}}|
\bar{c}\bar{d}\rangle_a$. In the PAC-CDFT, the latter three types of matrix elements vanish 
because of the spatial symmetries fulfilled by the nuclear density distribution. As a result, 
only the matrix elements $\langle a\bar{b}|\hat{V}_{\mathrm{pair}}|c\bar{d}\rangle_a$ need to 
be calculated.

The antisymmetric matrix elements of the pairing interaction in Eq. (\ref{Hpair-sp-basis}) can
be separated into a product of spin and coordinate space factors
\begin{equation}
  \langle a\bar{b}|\hat V_{\rm{pair}}|c\bar{d}\rangle_a =
    \Big\langle a\bar{b}\Big|W\frac{1}{2}(1-P^\sigma)\Big|c\bar{d}\Big\rangle_a.
\end{equation}
The operator $\frac{1}{2}(1-P^\sigma)$ projects onto the $S=0$ spin-singlet product state
\begin{equation}
  \frac{1}{2}(1-P^\sigma)|c\bar{d}\rangle=|c\bar{d}\rangle_{S=0}
    =\frac{1}{2} i^{n_y^c+n_y^d}(-1)^{n_y^c}\delta_{n_y^c+n_y^d
     +n_z^c+n_z^d,\rm{even}}[|\uparrow\downarrow\rangle-|\downarrow\uparrow\rangle]|
     n^c n^d\rangle,
\end{equation}
and the problem is reduced to the calculation of the spatial part of the matrix element
\begin{equation}
\begin{split}
  \langle a\bar{b}|\hat V^{\mathrm{pair}}|c\bar{d}\rangle_a
    =&\frac{1}{2}(-i)^{n_y^a+n_y^b}(-1)^{n_y^a}\delta_{n_y^a+n_z^a+n_y^b+n_z^b,
      \mathrm{even}}\\
     &\times i^{n_y^c+n_y^d} (-1)^{n_y^c}\delta_{n_y^c+n_z^c+n_y^d+n_z^d,
      \mathrm{even}}\langle n^a n^b|W|n^c n^d\rangle.\\
\end{split}
\end{equation}

The following formalisms are similar as those in Ref. \cite{nikvsic20103d}. The spatial
part of the matrix element
\begin{equation}
  \langle n^a n^b|W|n^c n^d\rangle\equiv2\int \mathrm{d}\bm{r}_1
    \mathrm{d}\bm{r}_2\mathrm{d}\bm{r}'_1\mathrm{d}\bm{r}'_2~\phi_{n_a}(\bm{r}_1)
    \phi_{n_b}(\bm{r}_2)W(\bm{r}_1,\bm{r}_2;\bm{r}_1',\bm{r}_2')
    \phi_{n_c}(\bm{r}_1')\phi_{n_d}(\bm{r}_2'),
\end{equation}
can be decomposed into three Cartesian components,
\begin{equation}
  \langle n^a n^b|W|n^c n^d\rangle=2GW_xW_yW_z.
\end{equation}

Here the detailed derivation of the $x$ component is given
\begin{equation}
  W_x=\int \mathrm{d}x_1\mathrm{d}x_2\mathrm{d}x_1'\mathrm{d}x_2'~\phi_{n^a_x}(x_1,b_x)
      \phi_{n^b_x}(x_2,b_x)\delta(X-X')P(x)P(x')\phi_{n^c_x}(x'_1,b_x)\phi_{n^d_x}(x'_2,b_x).
\end{equation}
By transforming to the center-of-mass and relative coordinates, and making use of the
1D Talmi-Moshinsky transformation, the integrals over the center-of-mass coordinates
$X$ and $X'$ are solved analytically, and one can find
\begin{equation}
\begin{split}
  W_x=&\sum_{N_x}M^{n_xN_x}_{n_x^a n_x^b}
    I_{n_x}(b_x)M^{n'_xN'_x}_{n_x^c n_x^d}I_{n'_x}(b_x),
\end{split}
\end{equation}
where the selection rules
\begin{equation}
  n_x^a+n_x^b=n_x+N_x,~~~~~n_x^c+n_x^d=n'_x+N'_x,
\end{equation}
have been used to eliminate the sums over $n_x$ and $n'_x$.

The $M^{n_xN_x}_{n_x^a n_x^b}$ denotes the 1D Talmi-Moshinsky brackets
\begin{equation}\label{T-M-bracket}
\begin{split}
  M^{n_xN_x}_{n_x^a n_x^b}=&\sqrt{\frac{n_x^a!n_x^b!}{n_x!N_x!}}\sqrt{\frac{1}
   {2^{N_x+n_x}}}\delta_{n_x^a+n_x^b,n_x+N_x}\sum_m(-1)^{n_x+m}
   \begin{pmatrix}
    N_x\\
    N_x-n_x^a+m
   \end{pmatrix}
   \begin{pmatrix}
    n_x\\
    m
   \end{pmatrix}.
\end{split}
\end{equation}

The $I_{n_x}(b_x)$ reads
\begin{equation}
  I_{n_x}(b_x)=\int\mathrm{d}x~\phi_{n_x}(x,b_{x})P(x).
\end{equation}
By making use of the generating function for the harmonic oscillator wave functions
\cite{nikvsic2014dirhb}, one can get
\begin{equation}\label{I-nx-integral}
  I_{n_x}(b_x)=\pi^{-1/4}\sqrt{\frac{b_x}{b_x^2+a^2}}
    \left(\frac{a^2-b_x^2}{a^2+b_x^2}\right)^{n_x/2}\sqrt{\frac{n_x!}{2^{n_x}}}
    \frac{1}{(n_x/2)!}\delta_{n_x,\mathrm{even}}.
\end{equation}

To summarize, finally, the antisymmetric matrix element of the pairing interaction
in Eq. (\ref{Hpair-sp-basis}) is
\begin{equation}
  \langle a\bar{b}|V^{\mathrm{pair}}|c\bar{d}\rangle=
    G\sum_{N_xN_yN_z}V_{a\bar{b}}^{N_xN_yN_z*}V_{c\bar{d}}^{N_xN_yN_z},
\end{equation}
which can be represented as a sum of separable terms in a 3DHO basis, with the
single-particle matrix elements
\begin{equation}
  V_{a\bar{b}}^{N_xN_yN_z}= i^{n_y^a+n_y^b}(-1)^{n_y^a}\delta_{n_y^a+n_z^a+n_y^b+n_z^b,
    \mathrm{even}}V_{n^a_xn^b_x}^{N_x}(b_x)V_{n^a_yn^b_y}^{N_y}(b_y)
    V_{n^a_zn^b_z}^{N_z}(b_z).
\end{equation}
The factors $V_{n^a_xn^b_x}^{N_x}(b_x)$ are given by
\begin{equation}
  V_{n^a_xn^b_x}^{N_x}(b_x)= M_{n^a_xn^b_x}^{n_xN_x}I_{n_x}(b_x)~~~
    \mathrm{with}~~n_x=n^a_x+n^b_x-N_x.
\end{equation}
The Talmi-Moshinsky brackets $M_{n^a_xn^b_x}^{n_xN_x}$ are defined in Eq.~
(\ref{T-M-bracket}), and the integrals $I_{n_x}(b_x)$ are given in Eq.~
(\ref{I-nx-integral}).
\end{appendix}


\end{CJK*}
\end{document}